# Growth of R-123 Phase Single Crystal Whiskers

Masanori NAGAO[1,2,*], Mitsunori SATO[2], Yukitake TACHIKI[2], Kinya MIYAGAWA[2], Masaki TANAKA[2], Hiroshi MAEDA[3], Kyung Sung YUN[1], Yoshihiko TAKANO[1] and Takeshi HATANO[1]

[1] National Institute for Materials Science, 1-2-1 Sengen, Tsukuba 305-0047, Japan
[2] Kitami Institute of Technology, 165 Koen-cho, Kitami 090-8507, Japan
[3] National High Magnetic Field Laboratory, Florida State University, Tallahassee, FL 32310, U.S.A.

Single-crystal whiskers of $R_1Ba_2Cu_3O_x$ (R-123, R= La, Nd, Sm, Eu, Gd, Dy, Ho, Er, Tm, Yb and Lu) phase have been successfully grown by the Te- and Ca-doping method. The whiskers were grown from precursor pellets at just below their partial melting (peritectic) temperatures. The nominal composition of the R-123 whiskers is $R_{1+u}Ba_{2+v}Ca_wCu_3O_x$ (u+v+w=0, w>0) with R and/or Ba sites being substituted by Ca. However, the amount of Te was less than the analytical limit. The critical temperatures of the R-123 whiskers were around 80 K, and among these whiskers those with larger R ionic radii such as Dy, Gd, Eu and Sm require post-annealing in an oxygen atmosphere.

(To be published in Jpn. J. Appl. Phys.)

Corresponding author
Name: Masanori Nagao
E-mail address: NAGAO.Masanori@nims.go.jp
Present address: National Institute for Materials Science, Tsukuba 305-0047, Japan.

------------------------------------------------------------------------

Since the discovery of high-$T_c$ cuprate superconductors such as Bi-based (Bi-2201, Bi-2212 and Bi-2223) [1] and $R_1Ba_2Cu_3O_x$ (R-123, R = Rare earth element) [2], much effort has been focused on the growth of single crystals, both for fundamental research and for electronic applications. The high-$T_c$ cuprate superconductors have stacked Josephson junctions along the $c$-axes of their crystal structures, called intrinsic Josephson junctions, which makes it possible to apply these superconductors in electronic device applications in the high-frequency terahertz (THz) band.[3,4] Therefore, high-quality single crystals play a key role in the development of the THz band electronic devices. The focus of recent studies has been on single crystal whiskers with nearly perfect crystallinity. Previously, we have grown single crystal whiskers of high quality Bi-2201, Bi-2212, and Bi-2223 from Te- and/or Ca-doped precursor pellets[5-7]. For the Bi-2212 and Bi-2223 single crystal whiskers, we have observed nearly perfect crystallinity of the whiskers and homogeneous multibranched structures in their current-voltage (*I-V*) characteristics along the *c*-axis, which exhibit intrinsic Josephson effects. In the previous work, we succeeded in growing single crystal whiskers of Y-123 ($Y_1Ba_2Cu_3O_x$) from Te- and Ca-doped $Y_1Ba_2Cu_3O_x$ precursor pellets with a nominal composition of $Y_1Ba_2Cu_3Te_{0.5}Ca_{1.0}O_x$[8]. However, there have been no reports of the successful fabrication of R-123 single crystal whiskers. Due to the systematic variation in the crystal structures and superconducting properties of the Y-123 and R-123 (R = rare earth element) series, it is important to examine R-123 single crystal whiskers from the viewpoints of both the fundamental understanding of the growth mechanism and electronic applications. The large variation in the ionic sizes of various rare earth ions (0.86 -1.05Å) is expected to be reflected in the growth conditions and the superconducting properties of series R-123 whiskers. Furthermore, the mutual substitution of constituent $R^{3+}$, $Ba^{2+}$ and $Ca^{2+}$ ions in the R-123 crystal lattice is expected to appear, which results in the change of $T_c$. With respect to R-123 bulks, it has been reported that the $R^{3+}$ ions with large radius (La, Nd, Sm, Eu and Gd) tend to substitute for Ba sites, causing a significant decrease in $T_c$[9]. Also, $Ca^{2+}$ ions substitute for the R sites in the case of smaller $R^{3+}$ ions such as Y[10] and Er[11].

In this paper, we report on the successful growth of single crystal whiskers of R-123 with $R^{3+}$ ions from La to Lu except for Pr, Ce and Tb. The effect of the $R^{3+}$ ionic radius $r_i$ on the growth condition, particularly on the holding temperature $T_{max}$ of the heat treatment process, is studied. We estimate the extent of substitutions of constituent $R^{3+}$, $Ba^{2+}$ and $Ca^{2+}$- ions and show the relation of $r_i$ to the chemical composition and hence to $T_c$ of the whiskers.

Precursor powders with a nominal composition of $R_1Ba_2Cu_3Ca_1Te_{0.5}O_x$, with R = La, Nd, Sm, Eu, Gd, Dy, Ho, Er, Tm, Yb, and Lu, were synthesized by the solid-state reaction of individual component oxides and carbonates. The powders of $R_2O_3$, $BaCO_3$, $CuO$, $TeO_2$ and $CaCO_3$, all with 99.9% purity, were thoroughly mixed and calcined in air at 800-900°C for 10 h with three intermediate grindings. The calcined powders were pressed into pellets of 10 mm in diameter and 2-3 mm in thickness. The precursor pellets were held at $T_{max}$ for 5 h in 20% $O_2$ (air) or 100% $O_2$, followed by slow cooling at a rate of 1°C /h down to 900°C. The samples were then furnace-cooled from 900°C to room temperature. Subsequently, several R-123 whiskers with R of Y, Dy, Gd and Sm were post-annealed at 300°C in an oxygen atmosphere. The length of the R-123 whiskers was observed to increase with the increase in the holding duration at $T_{max}$ and reached a maximum length in 5 h. The growth length is also increased with the decrease of the cooling rate from $T_{max}$ to 900°C, to less than 1°C /h . In this case, however, the tips of the whiskers exhibit a curvature and their surfaces become rough, leading to a deterioration in crystallinity. The whiskers were evaluated by X-ray diffraction (XRD), energy dispersive spectroscopic analysis (EDS), electron probe microanalysis (EPMA), and magnetization measurements using a superconducting quantum interference device (SQUID) magnetometer.

Figure 1 shows a scanning electron microscopy (SEM) image of a grown Dy-123 phase whisker. The whiskers originate from the inside of the pellet, and the whiskers exhibit a flat surface. With a width of around 20μm. Figure 2 shows an XRD pattern of the flat surface (shown as the a-b plane in Fig.1) of the Dy-123 whisker. The presence of sharp diffraction lines of only the (00$l$) indices of the Dy-123 (R-123) structure indicates that the flat surface is the a-b plane and the quality of the crystal is fine. All of the R-123 whiskers grown, except La, Nd, Yb and Lu, have a very flat surface, as shown in Fig.1, and are of 20-40μm width, 10-30μm thickness and 1-3mm length.

Figure 3 shows the holding temperature $T_{max}$ and the types of R-123 whiskers as a function of the $R^{3+}$ ion radius $r_i$, for the heating schedule where whiskers were found to grow under 20% $O_2$ atmosphere. The values for Y-123 whiskers[8] obtained in the previous work are also plotted in the figure for comparison. In the figure, the circles and triangle signify the average lengths of grown whiskers (●:2-3mm, ○:1-2mm, Δ:<1mm). The melting (peritectic) temperature of R-123 bulk materials without Ca substitution are shown for reference[12]. As the $R^{3+}$ ion radius $r_i$ increases, the optimum $T_{max}$ for whisker growth increases. $T_{max}$, particularly those for >1mm length whiskers, was found just below the melting (peritectic) temperature of the corresponding R-123 bulk materials. As shown in Fig.3, R-123 whiskers were found to grow for Yb, Tm, Er, Ho, Dy, Gd, Eu, Sm, Nd and La. Among these, reproducible results were obtained for R= Tm, Er, (Y), Ho, Dy, Gd,

Eu and Sm, whose $r_i$ values range between 0.88 and 0.96Å. Particularly for (Y), Dy and Gd, long whiskers with lengths of 2.0-3.0mm are grown. We also investigated whisker growth under a 100% $O_2$ atmosphere and found that the optimum $T_{max}$ is higher by 10-20°C than those of whiskers grown below 20% $O_2$ as shown in Fig.3. This is due to the increase in the melting (peritectic) temperature of R-123 with the increase of oxygen partial pressure. The results that both of the values of holding temperature $T_{max}$ obtained under 20% and 100% $O_2$ were close to the melting (peritectic) temperature of R-123 bulk materials without Ca show that the growth of whiskers is closely related to the partial melting of the precursors. It should be noted, in the case of Lu-123, that whiskers of 0.3 mm in length were grown in 100% $O_2$ while no whiskers were grown in 20% $O_2$.

EPMA and EDS analyses on the whiskers reveal the presence of Ca. The amount of Te was less than the analytical limit in the R-123 whiskers, as reported for Bi-based and Y-123 superconducting whiskers.[5-8] Figure 4 shows the values of u, v and w in $R_{1+u}Ba_{2+v}Ca_wCu_3O_x$ (u+v+w=0, w>0) of grown R-123 whiskers as a function of the $R^{3+}$ ion radius $r_i$, as determined by EPMA and EDS analyses. The following points can be noted on the basis of the data presented in Fig.4: (1) The excess R-content (u) varies with $r_i$ ranging from negative to positive. (2) In the $r_i$ range below Dy, the values of u, v, w show a slight variation, but are almost constant while all the values above $r_i$>0.92 show far above Gd. (3) In R-123 with smaller R-ions (below Dy), the Ca content (w) is nearly equal to the sum of the deficiencies in R content (-u) and Ba content (-v); in Y-123, -u = w and v = 0. (4) On the other hand, in R-123 with larger R-ions (above Gd), the value of u+w is equal to –v (u+w=–v or |u+w|=|v|).

From the above results the following conclusions can be drawn: $Ca^{2+}$ ions substitute for the R sites as well as for the Ba sites in R-123 whiskers in the smaller $R^{3+}$-ions of Tm, Er, Ho and Dy, except in Y-123 whiskers where Ca ions substitute only for Y sites. We consider that the difference in the substitutions of Ca between R-123 and Y-123 is due to the solubility limit of Ca into the site. It has been reported that for Ca content >0.22, the excess Ca ions substitute for the Ba sites in Y-123 bulk[10], and we consider that in the case of another R-123 which exceeds the solubility limit of Ca, the excess Ca ions substitute for the Ba sites. On the other hand, in the case of large $R^{3+}$ ions of Gd, Eu, Sm, Nd and La, the excess R ions, together with Ca ions, substitute for the Ba sites. The amount of Ba site substitutions is increased with increasing $r_i$, according to previous research on bulk samples.[9] As the $R^{3+}$-ions radius $r_i$ becomes larger, the substitutions of small radius ions such as $R^{3+}$ or $Ca^{2+}$ for the Ba site may be required in order to maintain the R-123 crystal structure since the $Ba^{2+}$ ion radius is larger than those of $R^{3+}$ and $Ca^{2+}$. In Eu-123 whiskers, obtained values of u and v were significantly deviated from the inclination of compositions (values of u and v) of other R-123 ($r_i$) whiskers. It has been reported that Eu includes a fraction of larger $Eu^{2+}$ ions along with the $Eu^{3+}$ ions. The mix valence might be taken into account to clarify this deviation. The growth of Lu-123 whiskers was confirmed by XRD, but the composition of the whisker could not be determined using EDS.

Figure 5 shows the critical temperature $T_c$ of as-grown and post-annealed R-123 whiskers as a function of the $R^{3+}$ ion radius $r_i$, as determined from magnetization-temperature curves measured using a SQUID magnetometer. In the figure, a solid line shows the $T_c$ values obtained for R-123 bulk materials without Ca substitution. The post-annealing of whiskers is carried out at 300°C for 50 h in a 100% $O_2$ atmosphere. For the as-grown whiskers with smaller $R^{3+}$-ions (below Dy and Y), in which Ca ions substitute for the R site or both R and Ba sites, $T_c$ is almost constant at about 80 K. The value of $T_c$ of 80 K for the whiskers is lower by about 10 K than that for bulk without Ca. A similar slight decrease in $T_c$ has been reported for Y-123 and Er-123 bulk material in which Ca ions substitute for R-site.[10, 11] On the other hand, for the whiskers with larger $R^{3+}$-ions (above Gd), in which Ba deficiency (-v) = excess R (u) + Ca doping (w), as-grown $T_c$ decreases significantly to 35-50 K. The $T_c$ values for Dy-, Gd- and Sm-123 whiskers recover up to around 80 K by post-annealing in an oxygen atmosphere. Therefore, the reduced $T_c$ for as-grown whiskers is mainly due to the oxygen deficiency which results in the deviation of carrier concentration from the optimum, as pointed out with respect to R-123 bulk[9, 10]. Why does oxygen deficiency appear for the whiskers in which the $R^{3+}$-ions substitute for the $Ba^{2+}$ site? The diffusion pass of oxygen in R-123 is considered to be the Cu-O chain layer because the diffusion rate along the c-axis is more than five orders smaller than that in the a-b plane[13]. More precisely, one of the oxygen sites is totally vacant in the Cu-O chain layer and is located parallel to the Cu-O chains. Since the oxygen diffusion pass (CuO chain layer) is sandwiched by the Ba layers, the possible explanations are (1) that the substituted $R^{3+}$ ions may trap oxygen and suppress the diffusion rate of oxygen in the whiskers during the cooling, and (2) that there exists a barrier when x of $R_{1+u}Ba_{2+v}Ca_wCu_3O_x$ exceeds 7 to compensate for the hole deficiency caused by the presence of excess R. In the case of Y-123 whiskers, no change of $T_c$ is found by with post-annealing. Therefore, we can conclude that values of $T_c$ of the R-123 whiskers are around 80 K due to Ca substitution of around 15%. The question arises why $T_c$ is constant for both u<1(or =1) (smaller $r_i$) and u>1 (Gd and Sm, larger $r_i$) R-123 whiskers. In the latter, the R ions substitute for the Ba site, such as $R_{1+x}Ba_{2-x}Cu_3O_y$, which would result in the decrease in hole concentration and hence in the decrease in $T_c$[14]. In our Gd, Eu and Sm -123 whiskers, because the substitution of R for Ba site and the Ca doping take place simultaneously, the hole concentration could be optimized by the additional hole doping from the chain layer to the $CuO_2$ planes for R>1 whiskers by the increase in x of $R_{1+u}Ba_{2+v}Ca_wCu_3O_x$ beyond 7, which has been reported for Nd and La -123[15]. Further investigation is underway to clarify this phenomenon. The R-123 structures were confirmed in Nd-123 and La-123 whiskers with even larger $R^{3+}$ ion size by XRD, but the whiskers do not exhibit any superconductive transition down to 5 K by SQUID for the as-grown samples.

In summary, we have grown single crystal whiskers of R-123 phase with R = Lu Yb, Tm, Er, Ho, Dy, Gd, Eu, Sm, Nd, and La, by using Te- and Ca-doped $R_1Ba_2Cu_3Ca_1Te_{0.5}O_x$ precursor pellets. The R-123 whiskers were grown from precursor pellets at just below their peritectic temperatures of R-123 (partial melting). The R-123 whiskers of 1.0-3.0 mm length are grown with the $R^{3+}$ ion radius $r_i$ ranging from Tm to Sm. The whiskers have a small amount of Ca substituting for R and/or Ba sites, but Te is not detected. For as-grown whiskers with smaller $R^{3+}$-ions such as Dy, Ho, Er and Tm, $Ca^{2+}$ ions are substituted for both R- and Ba- sites, resulting in slightly reduced $T_c$ values of around 80 K. On the other hand, for the whiskers with larger $R^{3+}$-ions such as Gd, Eu, and Sm, R ions substituted for the Ba site while their $T_c$ was kept constant at 80 K. A post-annealing in an oxygen atmosphere is therefore required for these whiskers because the as-grown whiskers have significantly reduced $T_c$ values of 35-50 K.

**Figure captions**

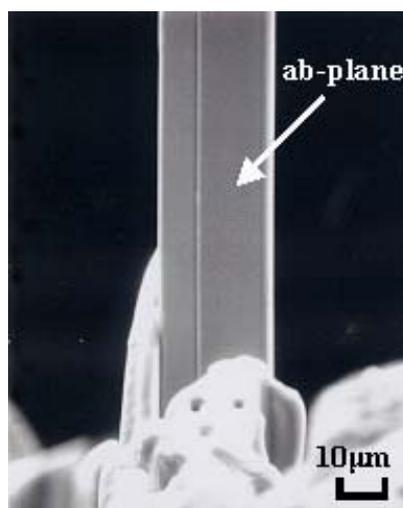

Fig. 1. SEM image of as-grown Dy-123 whisker.

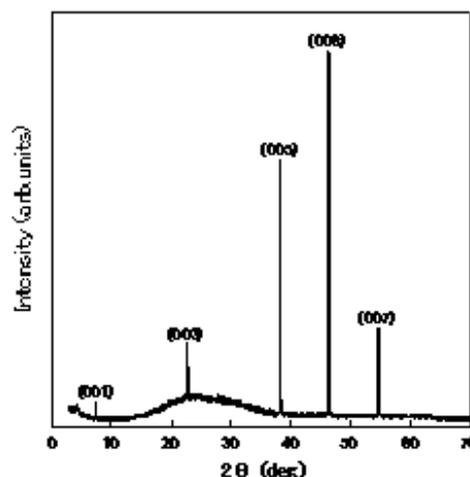

Fig. 2. XRD pattern of the wider surface of a Dy-123 whisker.

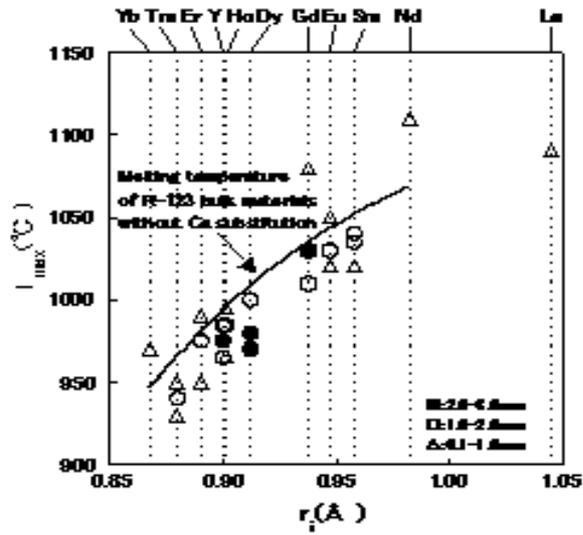

Fig. 3. Holding temperature $T_{max}$ and melting (peritectic) temperature of R-123 bulk materials[12] without Ca substitution as a function of $R^{3+}$-ion radius $r_i$ for R-123 in 20% $O_2$ atmospheres.

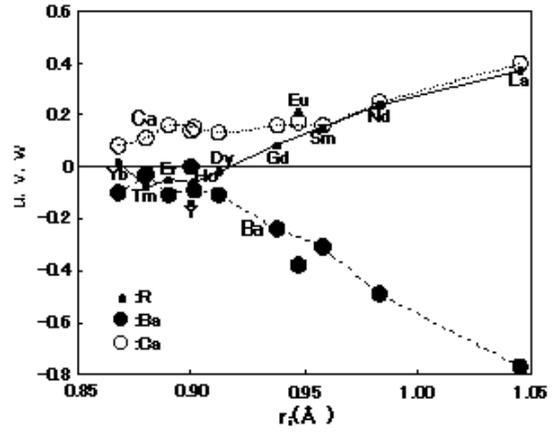

Fig. 4. Analyzed values of u, v, and w in $R_{1+u}Ba_{2+v}Ca_wCu_3O_x$ whiskers as a function of $R^{3+}$-ion radius $r_i$.

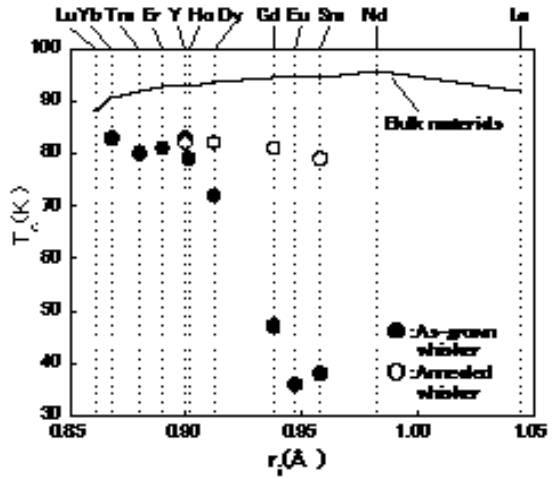

Fig. 5. Critical temperature $T_c$ as a function of $R^{3+}$ ion radius $r_i$ for as-grown and post-annealed R-123 whiskers and R-123 bulk materials without Ca substitution.